\documentclass  [11pt] {article}
\title {\bf Dimensional reduction in a model with infinitely many absorbing states
\vskip5mm}   
\author {{\sc Adam Lipowski}\\
\noalign{\vskip5mm}
{\it Department of Mathematics, Heriot-Watt University},\\ 
{\it EH14 4AS Edinburgh, United Kingdom}\\
and\\
{\it Department of Physics, A.~Mickiewicz University,}\\
{\it  61-614 Pozna\'{n}, Poland}\\
\noalign{\vskip5mm}}
\date {}
\newif\ifetykiety

\def\etykieta#1{\ifetykiety \par\marginpar{\centering[#1]} \fi}
\def\bibetykieta#1{\ifetykiety \marginpar{\renewcommand{\baselinestretch}{0.9}
                   \raggedright\small[#1]} \fi}

\newcommand {\beq} {\begin {equation}}
\newcommand {\eeq} [1] {\label {#1} \end {equation} \etykieta{#1}}
\newcommand {\beqn} {\begin {eqnarray}}
\newcommand {\eeqn} [1] {\label {#1} \end {eqnarray} \etykieta{#1}}
\catcode `\@=11
\def\@cite#1#2{#1\if@tempswa , #2\fi}
\catcode `\@=12
\newcommand {\cyt} [1] {$^{\mbox {\footnotesize \cite{#1})}}$}

\def\bib#1#2\par{\bibitem{#1} #2 \bibetykieta{#1}}
\newcommand {\fig} [1] {Fig.~\ref{#1}}
\newcounter {fig}
\newenvironment {figure_captions} {\newpage \thispagestyle {empty} \section*
{Figure captions} \begin {list} {\bf Fig.~\arabic{fig}:} {\usecounter{fig}
\settowidth{\labelwidth}{Fig.~999:} }}{\end{list}}
\def\elem#1#2\par{\item#2\label{#1}\etykieta{#1}}

\renewcommand {\baselinestretch} {1.0}
\renewcommand {\cyt} [1] {{\mbox [\cite{#1}]}}
\begin {document}
\maketitle
\begin {abstract}
Using Monte Carlo method we study a two-dimensional model with infinitely many absorbing
states.
Our estimation of the critical exponent $\beta\sim 0.273(5)$ suggests that the model
belongs to the (1+1) rather than to (2+1) directed-percolation universality class.
We also show that for a large class of absorbing states the dynamic Monte Carlo method leads to
spurious dynamical transitions.
\end {abstract}
Recently, nonequilibrium phase transitions have been very intensively studied.
To some extent this is motivated by their various potential applications ranging from
catalysis~\cyt{ZIFF}  to  epidemic processes~\cyt{GRASSTORE} to interfacial behaviour in random
media~\cyt{BARA}.
Another motivation is a desire to classify a rich behaviour of these systems into some
universality classes in a manner resembling a relatively complete classification of equilibrium
phase transitions.
The basic idea of such an approach is to identify a parameter which determines to which
universality class the model actually belongs to, as e.g., dimensionalities of the order parameter
and of the embeding space for the equilibrium phase transitions.
It was already suggested that for nonequilibrium models the corresponding parameter might be the
number of absorbing states and in particular models with a single absorbing
state should belong to the so-called directed-percolation (DP) universality class~\cyt{GRASSJEN}.
Up to now there is an ample numerical evidence in support of this conjecture~\cyt{DP}.

There are also some indications that certain other models might belong to the so-called 
parity-conserving universality class and most likely models with two absorbing states belong to
this class~\cyt{PARITY,JENSENDP}.
Little is known, however, about further classification of models with finitely many absorbing
states.

Models with infinitely many absorbing states constitute an important class.
Such models~\cyt{KOHLER,YALDRAM,JENSEN1D,JENSEN2D} arises mainly in the study of surface catalysis
but recently these models were related also with  the Self-Organized Criticality~\cyt{VESP} or
biological evolution~\cyt{LIPLOP}.
The critical behaviour of these models is very interesting.
In the one-dimensional case, steady-state exponents have (1+1) DP values but certain dynamical
exponents remain nonuniversal.
However, some scaling arguments supported by numerical results suggest that the sum of these
dynamical exponents is the same as in the (1+1) DP case~\cyt{MENDES}.

In the two-dimensional case, fewer results are available and they are less accurate.
For example, for the so-called dimer-dimer model Albano's estimations of the exponent $\beta\sim
0.5$ describing the behaviour of the order parameter are marginally consistent with
$\beta_{{\rm DP}}^2=0.592$ of the (2+1) DP~\cyt{ALBANO}.
Similar calculations for the dimer-trimer model~\cyt{KOHLER,JENSENPRL93} and for a certain variant
of a sandpile model~\cyt{VESP} also suggest the (2+1) DP universality class.
Moreover, there are some renormalization-group arguments that models with infinitely many absorbing
states should belong to DP universality class, at least with respect to the steady-state
properties~\cyt{MUNOZ}.
The result which does not seem consistent with the (2+1) DP, namely $\beta\sim 0.2-0.22$, was
reported by Yaldram et al. for a certain model of CO-NO catalytic reaction~\cyt{YALDRAM}.
However, it was suggested by Jensen that their calculations were not accurate enough and
additional calculations have shown that also in this case the model exhibits the (2+1) DP critical
behaviour~\cyt{JENSEN2D}.
Indeed, Jensen calculated a number of exponents and all of them were consistent with the DP
universality class.
However, to estimate the exponent $\beta$ and show that it is also consistent with the DP
universality class, he used the so-called dynamic Monte Carlo method combined with some scaling
relations, and such an approach is fundamentally different from the steady-state calculations by
Yaldram et al.
To refute Yaldram et al.'s result we would have be sure that indeed both approaches yield the same
results.
Although the dynamic Monte Carlo method is very frequently used and provides one of the most
accurate estimations of critical parameters, its extent and the very reason of applicability,
especially for models with infinitely-many absorbing states, is not, in our opinion, firmly
established and further examples either supporting or contradicting the usage of this method would
be very desirable.

In the present Rapid Communication we study the two-dimensional version of a certain model which
recently was introduced in the context of biological evolution~\cyt{LIPLOP}.
This model has infinitely many absorbing states and its critical behaviour in one 
dimension~\cyt{LIPLOP} is in agreement with other models of this kind.
Namely the critical exponent $\beta\sim 0.273$, describing the density of an active phase, is very
close to its (1+1) DP counterpart $\beta_{{\rm DP}}^1\sim 0.2765$~\cyt{JENSENDICK} and also the sum
of dynamical exponents $\eta$ and $\delta$ is universal (with respect to the choice of an absorbing 
state) and close to the DP value.
We show, however, that in two dimensions the model has a number of unexpected features.
First, our estimation of the exponent $\beta\sim 0.273(5)$ is clearly different than its (2+1) DP
counterpart $\beta_{{\rm DP}}^2\sim 0.592$~\cyt{GRASSDP}.
This value strongly suggests that in the steady state, due to a rather puzzling dimensional
reduction, the critical behaviour of the two-dimensional model is the same as of its
one-dimensional analogue~\cyt{CARDY}.
Moreover, we show that the applicability of the dynamic Monte Carlo method to this model is highly
questionable.
In particular, we show that there exists a large class of absorbing states for which this method
reproduces spurious dynamical transitions.
We expect that some other models with infinitely many absorbing states might also exhibit similar
behaviour.
Indeed, it was shown by Dickman that for certain two-dimensional model the estimation of critical
point using the dynamical Monte Carlo method depends on the choice of the initial configuration
and, in general, is different from the steady-state estimation~\cyt{DICKDEV}.
However, the reported difference was quite small (at most 4\%) and analysis of critical
behaviour is presumably affected by certain strong crossover effects~\cyt{GRASSDEV}.
Similar small differences were reported for yet another model with infinitely many absorbing
states~\cyt{MUNDEV}.
In the reported below results the difference between estimations of critical point using the
dynamical Monte Carlo method and steady-state method might be as large as 30\%. 

Our model is defined on a two-dimensional Cartesian lattice.
Omitting the biological interpretation, we assign a certain number $w_{i,j}$ to the bond
connecting sites $i$ and $j$ and such that $0<w_{i,j}<1$.
Introducing certain parameter $r$, we define the dynamics of our model as follows~\cyt{LIPLOP}:
(i) Choose a site $i$ at random.
(ii) Calculate $\omega=\sum_j w_{i,j}$, where summation is over all nearest neighbours $j$ of the
site $i$.
(iii) If $\omega > r$, then the chosen site $i$ is active and all bond variables $w_{i,j}$ are
replaced by the new ones, chosen randomly.
If $\omega<r$, the chosen site is nonactive and its bond variables remain unchanged.
It is obvious from the above rules that the model possesses infinitely many absorbing states.

Since a computational implementation of the above rules is straightforward, we present only the
results of our calculations.
First we measured the density $p$ of active sites (i.e., those with $\omega >r$) in the steady
state.
The initial configuration of bonds is chosen randomly.
Our results for various system sizes $L$ are shown in~\fig{f1}-\fig{f2}.
These results clearly indicate the phase transition separating the active ($p>0$) and absorbing
($p=0$) phases of the model.
Assuming that in the vicinity of the transition the density $p$ has a power-law singularity $p\sim
(r_{{\rm c}}-r)^{\beta}$ and using the least-square method, we estimate $r_{{\rm c}}=1.3867(5)$ and
$\beta=0.273(5)$.
These estimates are based on the results for $L=200$ but the estimation of $\beta$ based on
results for $L=100$ is very similar.
For $L=200$ we made runs of $10^5$ Monte Carlo steps neglecting for each $r$ data from the initial
$10^4$ Monte Carlo steps.
A Monte Carlo step is defined in a standard way, namely as a single, on average, update per site.

Our results show that as far as the steady-state properties are concerned, this model does not
belong to the (2+1) DP universality class.
They instead strongly suggest that both one- and two-dimensional versions have the same exponents
$\beta$ as the (1+1) DP.
At present we do not understand why such a dimensional reduction takes place.

In our opinion, Yaldram et al.'s model might have the same critical exponent $\beta$ as our model
(i.e., we suggest that their calculations were inaccurate but not as much as suggested by Jensen).
Why thus do Jensen's dynamic Monte Carlo calculations~\cyt{JENSEN2D} yield the (2+1) DP behaviour? 
Although at present we cannot locate the cause of these inconsistencies, below we show that the
dynamic Monte Carlo method requires serious reconsiderations when applied to models with infinitely
many absorbing states.

The idea of the dynamic Monte Carlo method is to set initially the model in one of the absorbing
states with a seed of the active phase in the centre of the system and study the subsequent
spreading of activity.
One expects that for a certain value of the control parameter of the model various characteristics
of spreading will exhibit a power-law scaling.
Moreover, there is a considerable numerical support, mainly from studying one-dimensional models, 
that such dynamical critical point coincides with the steady-state critical point independently on
the choice of an absorbing state.
We show, however, that for two-dimensional models this is not the case and the choice of the
absorbing state strongly affects the location and nature of the dynamical transition in the model.

First, let us consider a trivial example, where as an absorbing state we have chosen a state with
$w_{i,j}=w_0=0$ for all bonds except the bonds surrounding a certain site, which are chosen such
that this site is active.
It is easy to realize that for $r>1$ the activity cannot spread beyond this single site and the
system quickly returns to the absorbing state.
We do not present numerical data but we have checked that for $r<1$ the activity usually spreads
throughout the whole system, which indicates that the system is in the active phase.
With such a choice of an absorbing state, $r=1$ is the point which separates the active and
absorbing regimes of the model.
Although trivial, this is an example of an absorbing state for which the dynamical transition
($r=1$) does not coincide with the steady-state one ($r=1.3867$).
It is also easy to notice that any absorbing state with $w_0\in (0,\frac{r_{{\rm
c}}-1}{3})$ also yields spurious dynamical transition basically due to the same mechanism.

It is also natural to expect that such spurious transitions will appear even when
some inhomogeneous absorbing states are considered.
As an example let us consider an absorbing state where all bonds $w_{i,j}$ are chosen randomly
from the interval (0,0.1).
Using our previous analyses for homogeneous absorbing states one can see that for such a choice of
an absorbing state the dynamic transition must take place at $r_{{\rm c}}^{{\rm d}}\in (1,1.3)$ and
numerical simulations~\cyt{LIPPREP} show that indeed $r_{{\rm c}}^{{\rm d}}=1.21(1)$, which is well
below the steady-state critical point $r_{{\rm c}}=1.3867$.

In the above examples the bond variables were set to low values and thus the active phase
was strongly suppressed.
Setting bond variables to large values (but such that their sums for each site do not exceed the
threshold $r$), we can construct absorbing states where the absorbing phase is suppressed.
As an example let us consider the case of $w_0=0.25$.
Setting a central site in the active state, we measured the number of active sites $N(t)$ as a
function of time $t$ and the results in the logarithmic scale for various $r$ are shown
in~\fig{f3} (average is taken over all runs). 
The number of independent runs varied from 200 for $r=1.42$ to 10000 for $r=1.47$.
It is essential in this type of simulations to ensure that the propagating activity never reaches 
the border of the lattice; for example for $r=1.45$ we have to use $L=1500$.
It is also essential to keep a list of active sites since they constitute only a small fraction of
all sites.
One can see that $r=1.45(1)$ is a point which separates two regimes with asymptotically increasing
and decreasing number of active sites.
Why for $r_{{\rm c}}<r<1.45$, i.e., in the absorbing phase, might the activity spread for the
infinitely long time?
(Of course the activity can spread for the infinitely long time also for
$r<r_{{\rm c}}$ but that is justifiable since for such $r$ the system is in the active phase).
The reason for that is the unstable character of the absorbing state: large values of bond
variables considerably ease spreading of activity.
An example of such a propagating structure for $r=1.42$ is shown in~\fig{f4}.
We put a single site in the centre of the 1000x1000 lattice in the active state and recorded the
configuration after time $t=1000$.
One can see that the activity is restricted only to the gradually increasing boundary of such a
structure.
But this is not surprising: for $r=1.42$ the model is in the absorbing phase and activity in the
centre dies out after some transient time which is needed for the system to find a stable
absorbing state.
One can also say that once such a structure has spread to infinity, an unstable absorbing state is
transformed into the stable one.
Let us also notice that the asymptotic slopes in~\fig{f3} for $r \leq 1.45$ seem to be the same and
slightly larger than unity.
It suggests that structures like that shown in~\fig{f4} might be fractals with the fractal
dimension greater than unity and that this fractal dimension might be universal (i.e., independent
of $r$).
Similar propagating structures were observed also for other models with absorbing
states~\cyt{DICKDEV,GRASSDEV}.

We expect that $w_0=0.25$ is not the only value for which absorbing phase is suppressed and
spurious dynamical transition is obtained.
Similar results should be obtained also when a large-$w$ absorbing state contains some
inhomogeneity (i.e., bonds are random variables from a certain range).
But we also expect that there exists a large class of absorbing states, which are in some
sense in between these two extremal classes examined above and for which dynamic Monte Carlo method
will correctly locate the critical point.
However, it means that in this method the choice of the absorbing state is very important and
presumably it is very difficult to predict whether a given absorbing state will lead to a spurious
or true critical point.

In his calculations Jensen~\cyt{JENSEN2D} used so-called typical absorbing states, which most
likely correspond to our in-between states and which most likely correctly reproduce the transition
point and yield the (2+1) DP exponents.
We can only suggest that it is these extremal absorbing states which affect the steady-state
dynamics and are responsible for the change of the universality class of our model.
\begin {thebibliography} {00}

\bib {ZIFF} R.~M.~Ziff, E.~Gulari and Y.~Barshad, Phys.~Rev.~Lett.~{\bf 56}, 2553 (1986).

\bib{GRASSTORE} P.~Grassberger and A.~de la Torre, Ann.~Phys.~(N.~Y.) {\bf 122}, 373 (1979).

\bib {BARA} A.~L.~Barabasi, G.~Grinstein and M.~A.~Mu\~{n}oz, Phys.~Rev.~Lett.~{\bf 76}, 1481
(1996).

\bib {GRASSJEN} P.~Grassberger, Z.~Phys.~B {\bf 47}, 365 (1982).
H.~K.~Janssen, Z.~Phys.~B {\bf 42}, 151 (1981).

\bib {DP} J.~L.~Cardy and R.~L.~Sugar, J.~Phys.~A {\bf 13}, L423 (1980).
R.~C.~Brower, M.~A.~Furman and M.~Moshe, Phys.~Lett.~B {\bf 76}, 213 (1978).
R.~Durrett, {\it Lecture Notes in Particle Systems and Percolation} (Wadsworth, Pacific Grove, CA,
1988).

\bib {PARITY} H.~Hinrichsen, Phys.~Rev.~E {\bf 55}, 219 (1997).
H.~Takayasu and A.~Yu.~Tretyakov, Phys.~Rev.~Lett.~{\bf 68}, 3060 (1992).
A.~Lipowski, J.~Phys.~A {\bf 29}, L355 (1996).

\bib{JENSENDP} I.~Jensen, Phys.~Rev.~E {\bf 50}, 3623 (1994).

\bib{KOHLER} J.~K\"{o}hler and D.~ben-Avraham, J.~Phys.~A {\bf 24}, L261 (1991).

\bib {MUNOZ} M.~A.~Mu\~{n}oz, G.~Grinstein, R.~Dickman and R.~Livi, Phys.~Rev.~Lett.~{\bf 76}, 451
(1996).

\bib{YALDRAM} K.~Yaldram, K.~M.~Khan, N.~Ahmed and M.~A.~Khan, J.~Phys.~A {\bf 26}, L801 (1993).

\bib {JENSEN1D} I.~Jensen and R.~Dickman, Phys.~Rev.~E {\bf 48}, 1710 (1993).

\bib {JENSEN2D} I.~Jensen, Int.~J.~Mod.~Phys.~B {\bf 8}, 3299 (1994).
I.~Jensen, J.~Phys.~A {\bf 27}, L61 (1994).

\bib{VESP} A.~Vespignani, R.~Dickman, M.~A.~Mu\~{n}oz and S.~Zapperi, Phys.~Rev.~Lett.~{\bf 81},
5676 (1998).

\bib{LIPLOP} A.~Lipowski and M.~\L opata, Phys.~Rev.~E {\bf 60}, 1516 (1999).

\bib{MENDES} J.~F.~F.~Mendes, R.~Dickman, M.~Henkel and M.~C.~Marques, J.~Phys.~A {\bf 27}, 3019
(1994).

\bib{ALBANO} E.~V.~Albano, J.~Phys.~A {\bf 25}, 2557 (1992).

\bib{JENSENPRL93} for some discussion of these results see I.~Jensen, Phys.~Rev.~Lett.~{\bf 70},
1465 (1993).

\bib{JENSENDICK} I.~Jensen and R.~Dickman, J.~Stat.~Phys.~{\bf 71}, 89 (1993).

\bib{GRASSDP} P.~Grassberger, J.~Phys.~A {\bf 22}, 3673 (1989).

\bib {CARDY} Dimensional reduction is known to take place in e.g., some random systems: G.Parisi
and N.Sourlas, Phys.~Rev.~Lett.~{\bf 43}, 744 (1979).

\bib {DICKDEV} R.~Dickman, Phys.~Rev.~E {\bf 53}, 2223 (1996).

\bib {GRASSDEV} P.~Grassberger, H.~Chat\'{e} and G.~Rousseau, Phys.~Rev.~E {\bf 55}, 2488 (1997).

\bib {MUNDEV} M.~A.~Mu\~{n}oz, G.~Grinstein and R.~Dickman, J.~Stat.~Phys.~{\bf 91}, 541 (1998).

\bib{LIPPREP} A.~Lipowski, in preparation.

\end {thebibliography}
\begin {figure_captions}

\elem {f1} Density of active sites $p$ as a function of $r$ calculated for $L=50$ ($\triangle$),
100 
($\diamond$) and 200 (+).

\elem {f2} The logarithmic plot of density $p$ as a function of $(r_{{\rm c}}-r)$ for $L=200$ and
$r_{{\rm c}}=1.3867$.

\elem {f3} The logarithmic plot of the number of active sites $N(t)$ as a function of time $t$ for
(from the top to the bottom) $r=1.42, \ 1.43, \ 1.44, \ 1.45, \ 1.46$ and 1.47.

\elem {f4} Active sites (dots) propagating in the $w_0=0.25$ absorbing state for $r=1.42$.
A single active site was placed in the middle of the 1000x1000 lattice and the configuration
shown was recorded after time $t=1000$.

\end {figure_captions}
\end {document}
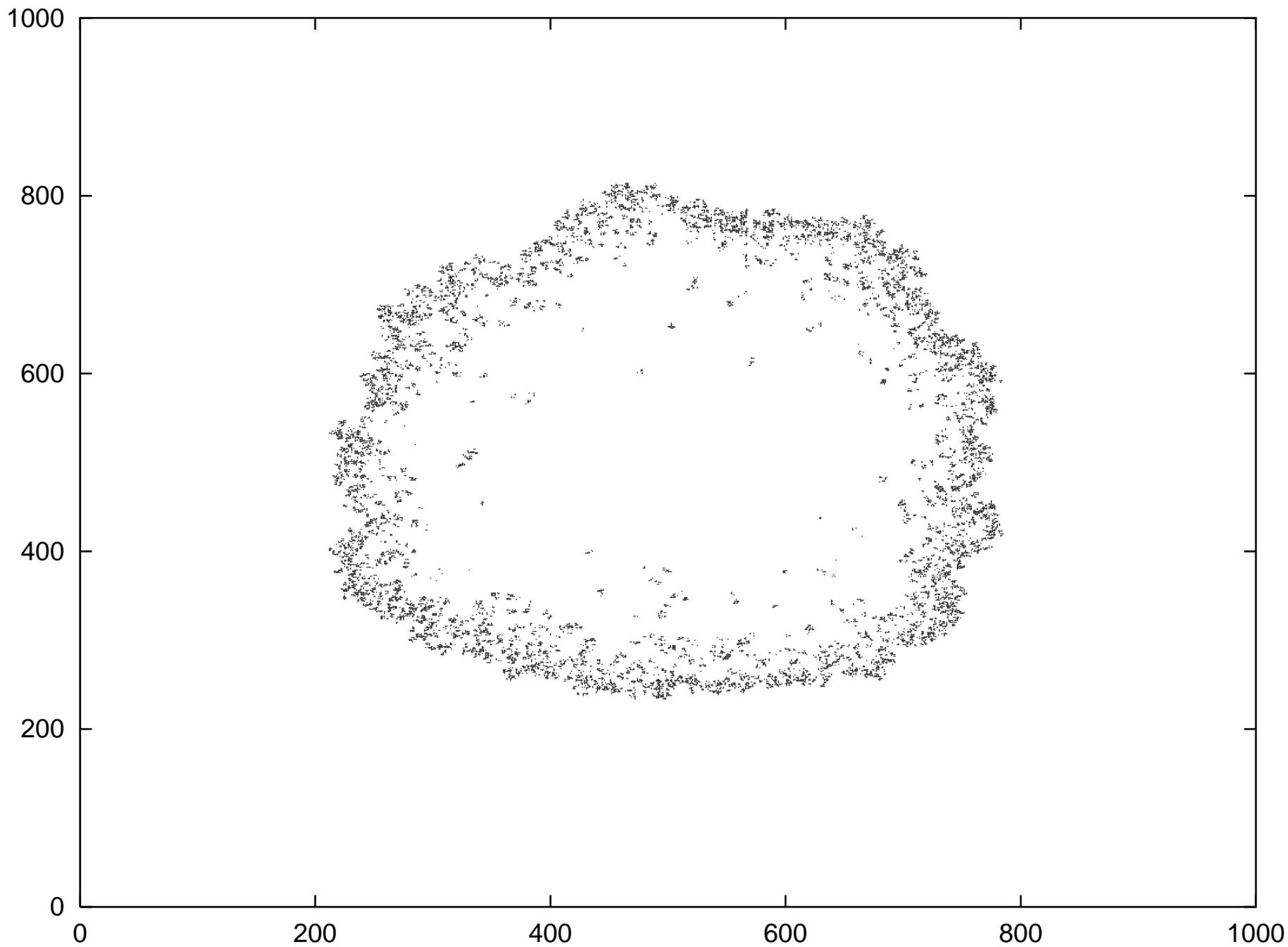
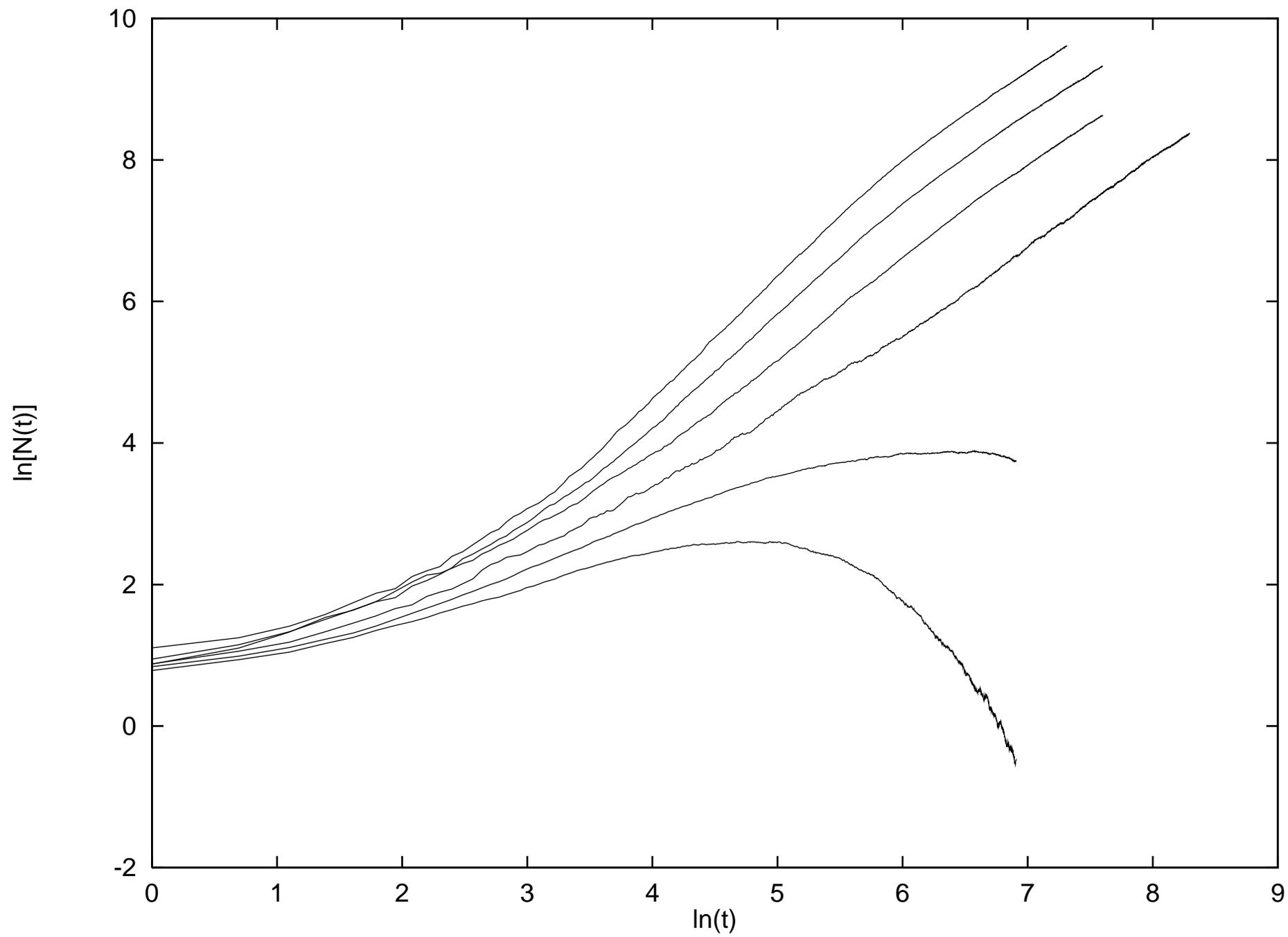
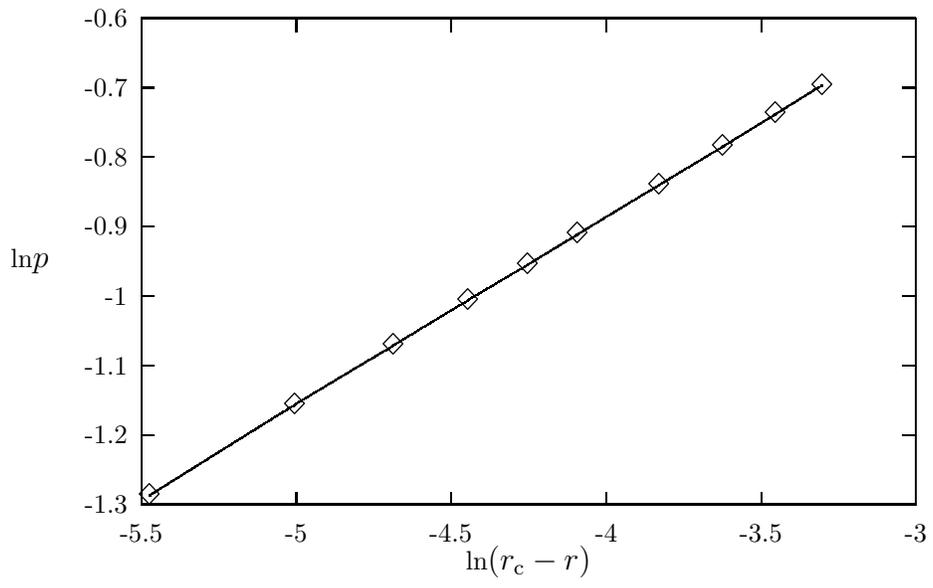
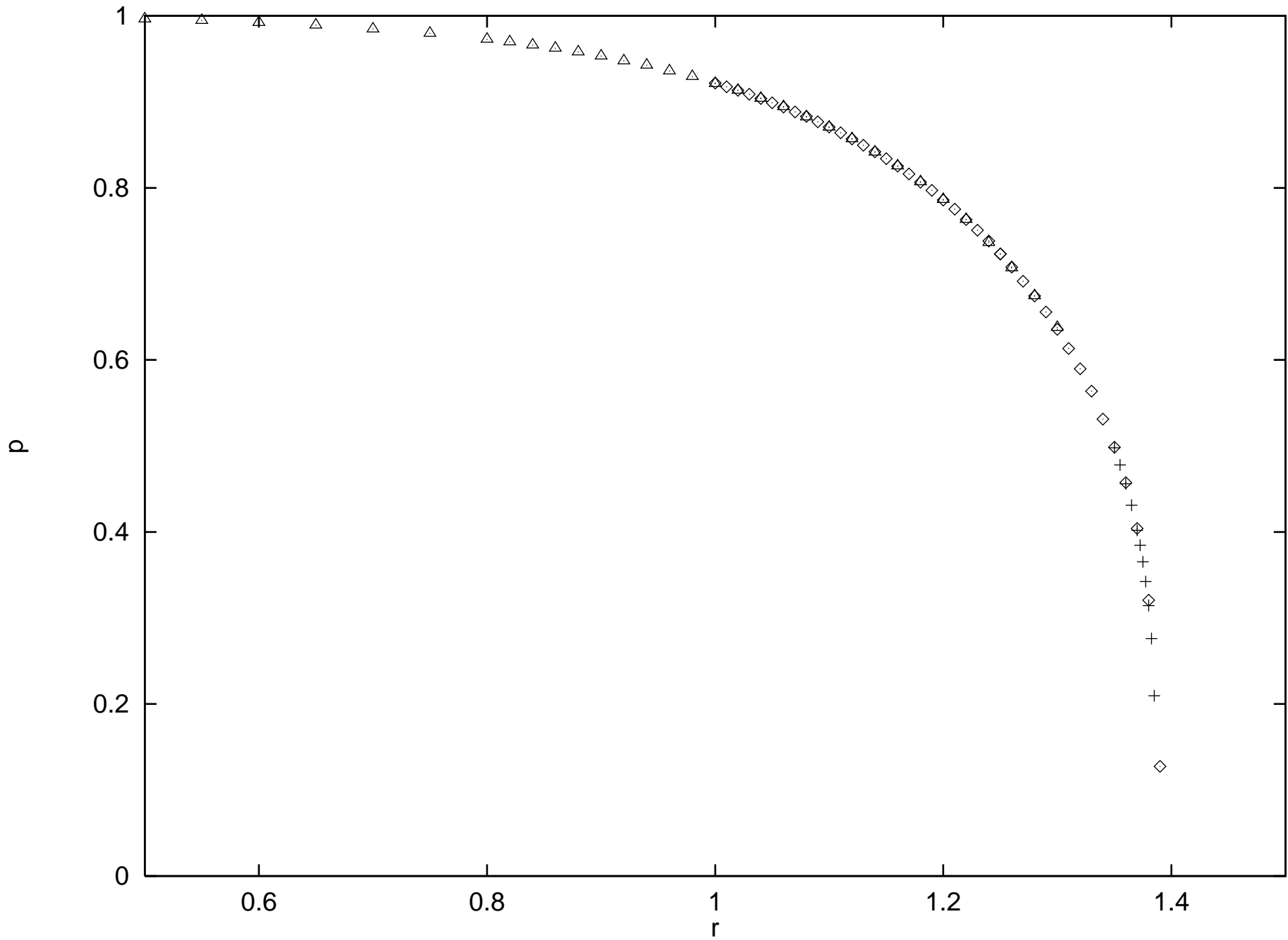
